\documentclass[11pt]{amsart}
\usepackage{amsmath, amssymb, graphicx}
\usepackage{mathrsfs}
\usepackage{txfonts} 
\usepackage{mathabx}
\newcommand{\no}[1]{#1}
\renewcommand{\no}[1]{}
\no{\usepackage{times}\usepackage[subscriptcorrection, slantedGreek, nofontinfo]{mtpro}
\renewcommand{\Delta}{\upDelta}}
\usepackage{color}

\date{\today}
\newtheorem{theorem}{Theorem}[section]
\newtheorem{proposition}{Proposition}[section]
\newtheorem{lemma}{Lemma}[section]

\newtheorem{corollary}{Corollary}[section]

\theoremstyle{remark}

\numberwithin{equation}{section}

\newcommand{\be}{\begin{equation}}
\newcommand{\ee}{\end{equation}}
\newcommand{\ba}{\begin{array}}
\newcommand{\ea}{\end{array}}

\newcommand{\bea}{\begin{eqnarray*}}
\newcommand{\eea}{\end{eqnarray*}}
\newcommand{\bean}{\begin{eqnarray}}
\newcommand{\eean}{\end{eqnarray}}

\title[Photoacoustic effect]{Mathematical modelization of 
the  Photoacoustic effect generated by the heating of metallic nanoparticles}

\author[Faouzi Triki]{Faouzi Triki$^\dag$}
\address{Laboratoire Jean Kuntzmann,  UMR CNRS 5224, 
Universit\'e  Grenoble-Alpes, 700 Avenue Centrale,
38401 Saint-Martin-d'H\`eres, France}

\email{Faouzi.Triki@univ-grenoble-alpes.fr}

\author[Margaux Vauthrin]{Margaux Vauthrin}
\address{Laboratoire Jean Kuntzmann,  UMR CNRS 5224, 
Universit\'e  Grenoble-Alpes, 700 Avenue Centrale,
38401 Saint-Martin-d'H\`eres, France}

\email{Margaux.Vauthrin@univ-grenoble-alpes.fr}

\date{\today}

\begin{document}

\begin{abstract}

This paper is devoted to the modelization of the  photoacoustic
effect generated by the electromagnetic heating of metallic nanoparticles
embedded in a biological tissue.  We first derive an asymptotic  models
for  the plasmonic resonances    and
the electromagnetic fields.
We then describe the acoustic 
generation created by the electromagnetic  heating of the nanoparticle.
Precisely, we derive the model equations that describes the
coupling between the temperature rise in the medium and the acoustic 
wave generation.  We obtain a direct relation between the acoustic waves
and the electromagnetic external sources.  
Finally, we solve the multiwave  inverse  problem that consists in the
recovery of the electric permittivity of the biological tissue from the 
measurements of the generated acoustic waves on the boundary 
of the sample.

\end{abstract}


\keywords{inverse problem, photo-acoustic, nanoparticle, plasmonic. \\ 
$^\dag$ This work is partially supported by Labex PERSYVAL-Lab (ANR-11-LABX-0025-01)}

\maketitle

\tableofcontents

\section{The Photoacoustic Model and main results}
\label{section1}
Photoacoustic imaging~\cite{BaUh-IP10, AmBoJuKa-SIAM10,  FSS, KiSc-SIAM13,
KuKu-HMMI10, BuMaHaPa-PRE07,  Scherzer-Book10, CoArBe-IP07, Wang-Book09,
AmBrJuWa-LNM12} is a
recent  hybrid imaging modality that couples electromagnetic 
waves with acoustic waves to achieve high-resolution imaging of
optical properties of heterogeneous media such as biological tissues.
Our objective in this paper  is  to derive a realistic 
complete mathematical model for the photoacoustic generation 
by a single nanoparticle embedded in a biological tissue. 
We introduce 
the mathematical  framework and give the main result in the
first section. In the second section we describe the mechanism of 
enhancement of light through the optical scattering properties of  
metallic
nanoparticles. The third section is devoted to the thermal
modelization of the part of the electromagnetic energy converted
into heat. We precisely derive  a theoretical model
for the  generation of acoustic waves by the thermal expansion
of the tissue around the metallic nanoparticles. The inverse
photoacoustic problem is solved asymptotically in section 4.
We finally give useful technical results in the appendix.  \\

We now give
a mathematical framework for the whole photoacoustic effect. 
Let $\Omega$ be a bounded
$C^2$ domain in $\mathbb R^2$.  The outward unit normal at $x$
to  $\partial \Omega $ is denoted by $\nu_{\Omega}(x)$. The domain
$\Omega$ is referred to as the biological sample that
we aim to image by the non-invasive photoacoustic modality.
Assume that $\Omega$ contains a single nanoparticule,  of 
the form $B_\alpha := z^\star+\alpha B$, where $B$ 
is a bounded, $C^2$ smooth domain containing the origin, 
$\alpha>0$ is a small  constant that represents the
size of the nanoparticle, and $z^\star$ is the position
of the nanoparticle. The first step  in photoacoustic imaging system
 is to illuminate the sample  by an electromagnetic wave produced by
 a laser 
 source. The time dependent, linear Maxwell's equations take the form 
\bea
\nabla \times {\bf E} &=& \hspace{-3mm} -\mu_0 
\frac{\partial }{ \partial t} {\bf H},
\\
\nabla \times {\bf H} &=&  \hspace{2mm} \varepsilon
 \frac{\partial }{ \partial t}
{\bf E},
\eea
where ${\bf E} $ and ${\bf H}$ are the total electric field and 
the total magnetic field respectively. The coefficients 
$\varepsilon$ and $\mu$ are the electric permittivity and magnetic 
permeability of the sample. The
magnetic permeability   is assumed to be constant
equals to $\mu_0$ the permeability of the free space, while the 
electric permittivity is given by  \\
 \bea
\varepsilon(x) \;=\; \left\{\ba{llcc} 
\varepsilon_s(x) \quad &\textrm{for   }  x\in\mathbb R^2 \setminus 
\overline{B_\alpha},\\
\varepsilon_m \quad &\textrm{for   } 
x\in B_\alpha,
 \ea \right.
\eea

where $\varepsilon_m$ is the permittivity
of the metal that will be specified later, and $\varepsilon_s(x)$
is the permittivity of the sample that is assumed of class $C^2$
and is constant equal to $\varepsilon_0>0$, the permittivity of the 
free space, outside $\Omega$. We assume throughout that 
$0<c_0<\Re(\varepsilon_s(x)) < \Re(\varepsilon_m)$, for all $x\in
\Omega$,  $\Im(\varepsilon_s(x))$ belongs $ C_0^2(\Omega)$
and satisfies  $\Im(\varepsilon_s (z^\star))>c_0$.  The imaginary 
part of the electric permittivity $\Im(\varepsilon_s(x))$,
 is related to the absorption of the electromagnetic energy,
  and provides a good description of the state of the biological
  tissue.  Our objective in this paper is to recover this 
parameter around the nanoparticles.    \\ 

 We assume that 
during the illumination of the sample a part of the electromagnetic
energy is dissipated by absorption inside the biological tissue and
inside the nanoparticle.  The absorption of the electromagnetic
energy by the biological tissue is transformed into heat and leads
 through the thermo-elastic 
expansion of the tissue to the generation of an 
acoustic pressure $p(x,t)$ that propagates to the detectors on
the boundary $\partial \Omega$. The measurements of 
$p(x,t)$  on the boundary allow the reconstruction of the 
absorption and diffusion coefficients  in the conventional plused 
photoacoustic imaging system.  In practice, it has been observed in various
experiments  that the imaging depth, i.e. the maximal depth of the sample at 
which features can be resolved at expected resolution,  is still fairly limited, usually 
on the order of millimeters. This is mainly due to the limitation on 
the penetration 
ability of the electromagnetic waves in the tissue:
 optical signals are attenuated significantly by absorption and scattering.
In~\cite{ChoulliTriki}, the authors showed that the resolution is
proportional to the magnitude of the laser fluence in the
sample, and recently in \cite{RenTriki} the mechanism of depth resolution 
is mathematically  investigated. 
 Metallic nanoparticles are very attractive as photoacoustic
contrast agents because of their large capacity to absorb light and
convert it to heat and their spectral selectivity.  When they are
illuminated at their plasmonic resonances their absorption of light 
is amplified and their temperature increases significantly leading to
various phenomena including heating the surrounding media. For example
in Hyperthermia therapy for cancer treatments one seeks to destroy
tumors through heating metallic nanoparticles \cite{pearce}. In the 
context of photoacoustic imaging the heat of the surrounding
biological tissue will generate a strong acoustic pressure 
wave $p(x,t)$ that can also be detected on the boundary $\partial
\Omega$. The principal idea for the use of metallic nanoparticles in 
photoacoustic imaging is that one can insert  them at any
position inside the sample and obtain strong acoustic sources
inside the sample. This will overcome the problem
of the limitation in the penetration resolution
depth of the conventional photoacoustic imaging modality 
based on the illumination of only the biological tissue. There
are already several related  results in the physicists 
community \cite{CFAE, bossy}.\\

 %We have the following main global stability estimate. 
 Our objective in this paper is to study the inverse problem to
recover  $\Im(\varepsilon_{s}(x))$ at $z^\star$ from measurements
of the pressure $p(x,t)$ on the boundary $\partial
\Omega$.\\

Assuming that $|\nabla {\bf H}(z^\star)|\not=0$,   $B$ is ball, and 
that $z^\star$ is known we derive the
the following global stability estimate. It shows how the errors in 
measurements can effect the reconstruction of the electric
permittivity at $z^\star$.  
%%%%%%%%%%%%%%%%%%%%%%%%%%%%%%%%%%%%%%%%%%%%%%%%%%%%%%%%%%
\begin{theorem} \label{main} Let $\tau_p > \tau_\Omega$ where
  $\tau_p= \sup_{x,y\in \Omega }|x-y|$. Let $p_a(x,t)$
  (resp. $p_b(x,t)$) the acoustic pressure generated by an
external electromagnetic source in a medium with electric
permittivity $\varepsilon_{s,a}(x)$ (resp. $\varepsilon_{s,b}(x)$). \\
  
Then, there exists a constant 
$C>0$ that does not depend on $\alpha$ and the boundary measurements
such that
 \bea
|\Im(\varepsilon_{s,a}(z^\star)) - 
\Im(\varepsilon_{s,b}(z^\star))| \leq 
C\left(\left\| \frac{\partial p_a}{\partial t}-
\frac{\partial p_b}{\partial t} \right\|_{L^2(\partial \Omega\times
(0,\tau_p))} + \| \nabla p_a -  \nabla p_b\|_{L^2(\partial \Omega\times
 (0,\tau_p))}\right)^{\frac{1}{4}} +O(\alpha).
\eea
\end{theorem}
%%%%%%%%%%%%%%%%%%%%%%%%%%%%%%%%%%%%%%%%%%%%%%%%%%%%%%%%%%%%
The proof of the theorem is given in section 4. It is based on 
asymptotic expansion of the electromagnetic fields when $\alpha$
tends to zero.  The coupling between the acoustic and  electromagnetic
waves allows us to retrieve the inner asymptotic expansion of
the electromagnetic fields in a small neiborhood
of $z^\star$ (Theorem \ref{resul}, and subsection 
\ref{innerexpansion}).  Since $\alpha$, the
size of the nanoparticle is small,  the stability estimate of H\"older 
type shows that the reconstruction of  $\Im(\varepsilon_s(z^\star))$  
from measurements of the pressure $p(x,t)$ on the boundary $\partial
\Omega$ is in fact  a well-posed inverse problem. In subsection
\ref{ssection3}
we derived the asymptotic expansion of the plasmonic resonances 
of the system nanoparticle and biological tissue. Later on
 in subsection \ref{innerexpansion},  we showed that  choosing the 
incident wave frequency close to the real part of a plasmonic
resonance enhance the photoacoustic signal measured on the 
boundary. Finally, the stability
result can be easily extended to cover the case where many well 
separated  nanoparticles are  embedded in the sample.

\section{Electromagnetic excitation}\label{section2}
The first syntheses of metallic small particles
date back to the 4th or 5th century BC where gold specimen were reported 
in China and Egypt. Their optical properties were used
for coloration of glass, ceramics, china and pottery
(see~\cite{Moores} and references therein).\\

 It is now well known 
that the interesting diffractive properties of  
these particles are linked  
to  resonances phenomena. In fact plasmon resonances may occur in metallic 
 particles  if the dielectric permittivity inside the particle is negative 
and the wavelength of the incident excitation is much larger than the 
dimension
of the particle.
For nanoscale metallic particles, these resonances occur in the
optical frequency range and they result in an extremely large
enhancement of 
the electromagnetic field 
near the boundary of the particles. This phenomena  has  applications
in many areas such as nanophotonics, nanolithography, near field
microscopy 
and biosensors. 
The desired resonance frequencies as well as the local fields
enhancement 
can be achieved 
by controlling the geometry of the metallic nanostructure.  \\

In mathematical point of view these resonances values
are the complex eigenvalues of Maxwell's equations
that only occur when the dielectric permittivity of the
nanoparticles is negative and the size of the nanoparticles 
 is less than the incident wavelength.  A formal asymptotic 
 expansion in \cite{MZ, MFZ} showed that  if the ratio between
 the incident wavelength and the size of the nanoparticle tends
 to zero the plasmonic resonances approches the eigenvalues 
 of the Neumann-Poincar\'e operator or the variational Poincar\'e 
 operator \cite{MZ, MFZ}.  In~\cite{BonnetierTriki}, 
the authors  have derived a rigorous justification
of the quasi-static approximation in harmonic frequency
regime~\cite{MZ, MFZ}. It is well known that the resonance
phenomena occur only in  transverse magnetic polarization (TM) polarization. 
Here we consider the time harmonic regime in (TM) polarization
  that is,  ${\bf E} = \Re(\mathbb E e^{i\omega t})$ and ${\bf H} =
   \Im(\mathbb H e^{i \omega t})$,
  where $\mathbb E =  (E(x_1,x_2), 0) $
and   ${\mathbb H} = (0,  0, H(x_1, x_2))$. \\

The total magnetic field  can be decomposed into two parts  
$H=  H_i + H_s$  where
$H_i$  and $H_s$  are respectively  the incident and scattered  waves.   \\

The homogeneous frequency-domain, linear Maxwell's equations,
in the transverse magnetic polarization (TM) and in absence of internal sources,
 take the form 
\begin{eqnarray} \label{mainequation1}
\nabla \cdot \left(\frac{1}{\varepsilon} \nabla H\right) +\omega^2 \mu_0 H &= 0&
\quad \textrm{in} \quad \mathbb R^2.
\end{eqnarray} 
with the  Sommerfeld radiation condition as $|x| \rightarrow +\infty$~\cite{Nedelec}:
\bean\label{mainequation2}
\frac{\partial H_s}{\partial |x|} -i\omega\sqrt{\varepsilon_0 \mu_0}H_s &=& O(\frac{1}{\sqrt{|x|}}).
\eean

Recall that the electric permittivity is given by  
\bea
\varepsilon(x) \;=\; \left\{\ba{llcc} 
\varepsilon_0 \quad &\textrm{for   }  x\in\mathbb R^2 \setminus 
\overline{\Omega},\\
\varepsilon_s(x) \quad &\textrm{for   }  x\in\Omega \setminus 
\overline{B_\alpha},\\
\varepsilon_m(\omega) \quad &\textrm{for   } 
x\in B_\alpha,
 \ea \right.
\eea

where  $ \varepsilon_0$ is the permittivity
of the free space.  The incident field  $H_i$ satisfies 

\begin{eqnarray*} 
\Delta H_i +\omega^2 \mu_0\varepsilon_0 H_i &= 0&
\quad \textrm{in} \quad \mathbb R^2.
\end{eqnarray*} 

The electric field $E$ can deduced directly from the 
magnetic field through  the  relation 
\bean\label{ef}
E(x) &=& \left(\ba{llcc}
\partial_{x_2} H(x)\\
-\partial_{x_1} H(x)
\ea\right). 
\eean
The  metal that fills the nanoparticle is assumed 
to be real and its
dielectric constant is  described by the Drude model:
\bean \label{Drude}
\varepsilon_m(\omega) &= & \varepsilon_0\left(\varepsilon_\infty -
\frac{\omega_P^2}{\omega^2+i\omega \Gamma}\right),
\eean
where $\varepsilon_\infty>0,~\omega_P>0$ and~$\Gamma>0$ are the metal  parameters 
that are usually
fitted utilizing experiment data~\cite{Moores}.  The dielectric constant $ \varepsilon_m $
depends on the frequency $\omega$, and so incident waves can cause a change
in the metal behavior.   Media having such a property are termed dispersive media.\\

The Drude model considered here describes well the optical properties of many metals within 
relatively wide frequency range.
 For example the function $ \varepsilon_m(\omega) $
with effective parameters: $\varepsilon_\infty = 9.84~eV$, $\omega_P = 9.096~ eV$,
$\Gamma = 0.072~eV$ for gold, and   $\varepsilon_\infty = 3.7~ eV$, $\omega_P = 8.9~ eV$,
$\Gamma = 0.021~ eV$ for silver reproduce quite well the experimental values of the 
dielectric constant in the frequency range $0.8~ eV$ to $4~ eV$ (see for instance~
\cite{Johnson-Christy}). 
%%%%%%%%%%%%%%%%%%%%%%%%%%%%%%%%%%%%%%%%%%%
\subsection{Plasmonic resonances}  \label{ssection3}
%%%%%%%%%%%%%%%%%%%%%%%%%%%%%%%%%%%%%%%%%%%

When the frequency lies in the upper half complex space, that is,  
$\Im (\omega)\geq 0$,
the system \eqref{mainequation1}  has a unique solution. 
The resolvent of the differential operator (\ref{mainequation1}) with
condition  
(\ref{mainequation2}) has a  meromorphic continuation in the lower 
complex plane. \\

The complex number $\omega$ is said to be a plasmonic 
resonant frequency of the nanoparticle
$B_\alpha$  if  there exists a  non-trivial solution $H$ to the
system~(\ref{mainequation1})-(\ref{mainequation2}) with zero 
incident wave.  \\

It is known that the set of scattering resonances $\{\omega_j\}$ of 
the above Helmholtz
equation in the absence of dispersion ($\varepsilon_m$ does not depend on $\omega$)  is 
discrete and symmetric in the complex plane 
about the imaginary axis.  Further,
it can be easily seen that all the resonant frequencies $\{\omega_j\}$  are in the lower half-space
$\Im \omega <0$. They can be found explicitly for a circular  or ellipsoid shape and are connected
in this case with the zeros of certain Bessel functions. More elaborate results assert that 
 for strictly convex shapes in dimension three the resonant frequencies accumulate
rapidly on the real axis as $|\Re \omega|\rightarrow \infty$~\cite{Popov}. \\
 It has been shown in dimension one that the scattering  resonances of a non-dispersive medium
 satisfy \cite{Harell, OstingWeinstein}   
\bea
\Im(\omega) \geq C_1 e^{-C_2|\Re(\omega )|^2},
\eea
where the constants $C_i, \; i=1,2$
only depend on $\varepsilon$ and  the size of
the domain.  \\
The imaginary part of a resonance gives the decay rate of the associated resonant states. 
Thus, resonances close to the real axis give information about long term behavior of waves.
 In particular, since the  work of Lax-Phillips \cite{LP} and Vainberg \cite{V}, resonance free 
 regions near the real axis have been used to understand decay of waves. Several works in 
 nano-optics have related the amplification and  enhancement of light to the behavior of the 
 imaginary part of the scattering resonances close to the real axis \cite{Ammari3, BT, BBT}.  \\

Alike the non-dispersive case,   the plasmonic resonances 
form a  set of discrete and isolated  complex values 
$\left(\omega_j(\alpha)\right)_j$. 
In \cite{BonnetierTriki}  the authors have derived the asymptotic expansion 
of the plasmonic resonant frequencies as $\alpha$ 
 tends to zero and  when the nanoparticle is surrounded
by a homogeneous medium with a constant electric permittivity.
We adapt  in the following paragraph their techniques to our problem and
derive the first term in the asymptotic expansion of the plasmonic resonances.  
We refer the reader to \cite{ADM, AKa, AKL, AMRZ} for recent and interesting 
mathematical results on plasmonic resonances for nanoparticles.\\

%%%%%%%%%%%%%%%%%%%%%%%%%%%%

Making the change of variables $x = z^\star+\alpha \xi$
in the spectral problem \eqref{mainequation1}, we get

\begin{eqnarray} \label{mainequation12}
\nabla \cdot \left(\frac{1}{ \tilde 
\varepsilon_\alpha} \nabla \widetilde H\right) + \alpha^2 \omega^2 \mu_0 \widetilde H &= 0&
\quad \textrm{in} \quad \mathbb R^2,
\end{eqnarray} 
with the radiation condition
\bean \label{mainequation22}
\frac{\partial {\widetilde H}}{\partial |\xi|} -i\alpha
 \omega \sqrt{\varepsilon_0 \mu_0}  \widetilde H &=& O(\frac{1}{\sqrt {|\xi|}}) 
 \quad \textrm{as} \quad  |\xi| \rightarrow +\infty,
\eean
where $ \widetilde H(\xi) = H( z^\star+\alpha \xi)$, and $\tilde \varepsilon_\alpha(\xi) =
\varepsilon(z^\star+\alpha \xi)$ is given by
\bea
\tilde \varepsilon_\alpha(\xi) \;=\; \left\{\ba{llcc} \varepsilon_0 \qquad &\textrm{for} 
&\xi \in\mathbb R^2\setminus 
\overline{\Omega_\alpha},\\
 \varepsilon_s(z^\star+\alpha \xi) \qquad &\textrm{for} 
&\xi \in\Omega_\alpha \setminus 
\overline{B},\\
\varepsilon_m(\omega)\qquad &\textrm{for} & \xi \in B.
 \ea \right.
\eea

Here  $\Omega_\alpha$ denotes $\left\{\frac{x-z^\star}{\alpha}; 
x\in \Omega \right\}$. It contains 
zero and tends to the whole space when $\alpha$ approaches  zero.  Similarly the piecewise 
smooth function $\tilde \varepsilon_\alpha(\xi) $ converges in $L^\infty_{loc}(\mathbb R^2)$ to
the piecewise constant function
\bea
\tilde \varepsilon_0(\xi) \;=\; \left\{ \ba{llcc}  \varepsilon_s(z^\star) \qquad &\textrm{for} 
&\xi \in\mathbb R^2\setminus \overline{B},\\
\varepsilon_m(\omega)\qquad &\textrm{for} & \xi \in B.
 \ea \right.
\eea

In the quasi-static regime  $\alpha\omega <<\omega<<1$, the
above spectral problem  formally converges, to the quasi-static 
spectral problem
\begin{eqnarray} \label{mainequation111}
\nabla \cdot \left(\frac{1}{ \tilde 
\varepsilon_0} \nabla \widetilde H_0\right)  &= 0&
\quad \textrm{in} \quad \mathbb R^2,
\end{eqnarray} 
where the field $\widetilde H_0(x)$ belongs to $W^{1,-1}_0(\mathbb R^2)$, where
 $$W^{1,-1}_0 (\mathbb R^2):= \left\{u\in H^1_{loc}(\mathbb R^2): \;\; u/(1+|\xi|^2)^{\frac{1}{2}}
 \ \ln(1+|\xi|^2)\in L^2(\mathbb R^2); \nabla u
\in L^2(\mathbb R^2); \;\; \lim_{|\xi| \to +\infty}u =0 \right\}. $$  
Next, we define the integral operator  $\mathcal T_0:  
W^{1,-1}_0 (\mathbb R^2) \rightarrow W^{1,-1}_0 (\mathbb R^2)$ by
\bea
\int_{\mathbb R^2 }\nabla \mathcal T_0 w \nabla v d\xi  =    \int_{B }
\nabla w \nabla v d \xi   \qquad \textrm{for  all    } v\in   W^{1,-1}_0 (\mathbb R^2).
\eea
We introduce the single  layer vector  space 
\bea
\mathfrak H := \{ u\in W^{1,-1}_0 (\mathbb R^2): \Delta u = 0 \;\; \textrm{in  }  B\cup 
 \mathbb R^2 \setminus \overline B;\;\;  u|_+ = u|_-
\;\; \textrm{on  }  \partial B  \}.
\eea

We deduce from \cite{BonnetierTriki} that the restriction 
of $\mathcal T_0$  to $\mathfrak H$ is a self-adjoint operator
of Fredholm type with index zero.  In fact $\frac{1}{2}I- \mathcal T_0$ is 
a compact  operator .  \\

Let us denote as $\left\{ \beta_j^\pm\right\}_{j \geq 1}$ the eigenvalues of $\mathcal T_0 : {\mathfrak H} \rightarrow {\mathfrak H}$, ordered  in the following way:
$$ 0 = \beta_1^- \leq \beta_2^- \leq ... \leq \beta_\infty^+=\frac{1}{2}, $$
and 
$$ \beta_\infty^+= \frac{1}{2} \leq... \leq \beta_2^+ \leq \beta_1^+ <1,$$
and satisfies  $\lim_{j\to +\infty }\beta_j^\pm =  \beta_\infty^\pm = \frac{1}{2} $. 
We deduce immediately  from the min-max principle for the compact, self-adjoint operator $\frac{1}{2}I- \mathcal T_0$ the  following characterization of the spectrum of $\mathcal T_0$ \cite{BonnetierTriki}.

%%%%%%%%%%%%%%%%%%%%%%%%%%%%%%%%%%%%%%%%%%%%%%%%%%%%%%%%%%%
\begin{proposition}\label{propminmax}
Let $\left\{ w_j^\pm \right\}_{j \geq 1}$ be the set of corresponding eigenfunctions of the operator $\mathcal T_0$,
associated to the eigenvalues $\left\{ \beta_j^\pm\right\}_{j\geq 1}$. The following equalities hold
$$\beta_j^{-} = \min\limits_{u \in {\mathfrak H} \atop u \perp w_1^-,...,w_{j-1}^-}{\frac{ \displaystyle{\int_D{\lvert \nabla u \lvert^2 \:dx}}}{ \displaystyle{\int_\Omega{\lvert \nabla u \lvert^2 \:dx}}}} = \max\limits_{F_j \subset {\mathfrak H} \atop \text{\rm dim}(F_j) = j-1}{\min\limits_{u \in F_j^\perp}{\frac{ \displaystyle{\int_D{\lvert \nabla u \lvert^2 \:dx}}}{ \displaystyle{\int_\Omega{\lvert \nabla u \lvert^2 \:dx}}}}}, $$
and
$$\beta_j^{+} = \max\limits_{u \in {\mathfrak H} \atop u \perp w_1^+,...,w_{j-1}^+}{\frac{ \displaystyle{\int_D{\lvert \nabla u \lvert^2 \:dx}}}{ \displaystyle{\int_\Omega{\lvert \nabla u \lvert^2 \:dx}}}} = \min\limits_{F_j\subset {\mathfrak H} \atop \text{\rm dim}(F_j) = j-1}{\max\limits_{u \in F_j^\perp}{\frac{ \displaystyle{\int_D{\lvert \nabla u \lvert^2 \:dx}}}{ \displaystyle{\int_\Omega{\lvert \nabla u \lvert^2 \:dx}}}}},$$
for all $j\geq 1$.
\end{proposition}
%%%%%%%%%%%%%%%%%%%%%%%%%%%%%%%%%%%%%%%%%%%%%%%%%%%%%%%%%%%%%%%%%%%%%%%%%%%

We define the quasi-static resonances 
$\left(\omega_{j}^\pm(0)\right)_{j\geq 1}$ of the 
spectral problem  \eqref{mainequation111} the complex roots 
of  the following
dispersion equations
\bean\label{dispersion}
\frac{\varepsilon_m(\omega)}{ \varepsilon_s(z^\star) }&=&k_j^\pm:=
\frac{ \beta^\pm_j}{ \beta_j^\pm-1}, \qquad 1 \leq j\leq \infty.
\eean

We first remark that  since $\beta_j^\pm $ belong
to $[0, 1)$ the values on the right side of the equality $k_j^\pm $ 
are negative reals. Thus  $\Re(\varepsilon_m(\omega))$,
 the real part of the electric permittivity,   at the 
quasi-static plasmonic resonances $\left(\omega_{j}^\pm(0)\right)_{j\geq 1}$
 takes  negative reals. This is exactly what one would expect in a such
situation, and the existence of the plasmonic resonances can not occur 
if  the material inside the nanoparticle is a  modest electric
permittivity that has always  a strictly positive real part.

 %%%%%%%%%%%%%%%%%%%%%%%%%%%%%%%%%%%%%%
\begin{lemma} \label{quasiresonance}
 The complex roots 
to the dispersion relation \eqref{dispersion}  are  explicitly given by
\bean
 -i\frac{\Gamma}{2}\pm\sqrt{\frac{\omega_p^2}{\varepsilon_\infty
    -k_j^\pm\varepsilon_s(z^\star)} -
\frac{\Gamma^2}{4}},
\eean
where $\sqrt{z}$ is the complex square root function defined on 
$\mathbb C\setminus i(0,\infty)$.

\end{lemma}
%%%%%%%%%%%%%%%%%%%%%%%%%%%%%%%%%%

We note that the quantities  $k_j$ and 
$ \varepsilon_\infty- 
4\frac{\omega_P^2}{\Gamma^2}$ only  depend  respectively on the 
shape of the particle and the nature of the metal that fills the particle. 
Based on this calculation we remark that the circular shape
has only four quasi-static resonances given by 
$-i\frac{\Gamma}{2}\pm\sqrt{\frac{\omega_p^2}{\varepsilon_\infty} -
\frac{\Gamma^2}{4}}$, and  
$-i\frac{\Gamma}{2}\pm\sqrt{\frac{\omega_p^2}{\varepsilon_\infty
    +\varepsilon_s(z^\star)} -
\frac{\Gamma^2}{4}}$. They satisfy respectively the dispersion equation
with $ k_1^-\,=0$, and $ k_\infty^\pm= -1$.  We remark that only the
resonances related to  $ k_\infty^\pm= -1$ depend on the  surrounding
media
electric permittivity $\varepsilon_s(z^\star)$ and may givel later
information on it. Finally, the 
eigenfunctions associated to $ k_1^-\,=0,$ 
are constant on the boundary $\partial B$. \\

We follow the same steps as in the proof of Theorem  2.1 in 
\cite{BonnetierTriki, AT, Ammari3}
and  prove the following asymptotic result. 
%%%%%%%%%%%%%
\begin{proposition}\label{asymp}
Let $\omega(0)$ be a quasi-static resonance with multiplicity  $m$.
Then there exist a constant $\alpha_0>0$ such that for $0<\alpha<\alpha_0$
there exist $m$ plasmonic resonances  
$\left(\omega_j(\alpha)\right)_{1\leq j\leq m}$
satisfying the following asymptotic expansion as $\alpha \to 0$:      
\bean
\frac{1}{m}\sum_{j=1}^{m} \omega_j(\alpha) = \omega(0)+ o(1).
\eean
\end{proposition}

Next, we derive the asymptotic expansion of the electromagnetic fields
when the size of the nano-particle tends to zero.  
%%%%%%%%%%%%%%%%%%%%%%%%%%%%
\subsection{Small volume expansion of the EM fields}  \label{ssection4}
Our strategy here is to use the tools  developed in \cite{Ammari2, AK}
and references therein to derive the leading terms in
the asymptotic expansion of electromagnetic
fields when  the volume of the nano-particle tends to zero.  Since the frequency
of the incident wave is real and thus far away from the complex
plasmonic resonances
we expect that the remaining terms of the asymptotic  expansion stay 
uniformly bounded. \\

Let $H_0= H_i+H_{0s}$,  be the total electric field in the  absence of the nanoparticle. 
It satisfies the
system 
\begin{eqnarray} \label{u1}
\nabla \cdot \left(\frac{1}{ \varepsilon_s} \nabla H_0\right) +\omega^2 \mu_0 H_0 &= 0&
\quad \textrm{in} \quad \mathbb R^2.
\end{eqnarray} 
with the  Sommerfeld radiation condition as $|x| \rightarrow +\infty$:
\bean \label{u2}
\frac{\partial H_{0s}}{\partial |x|} -i\omega\sqrt{\varepsilon_0 \mu_0}H_{0s} 
&=& O(\frac{1}{\sqrt{|x|}}).
\eean

Recall that in the quasi-static regime the   scattering resonances
are far away from the real axis.  Consequently  the  system \eqref{u1}-\eqref{u2} above has a 
unique solution $H$ for any given real frequency $\omega$. Hence the following Green function
$G(x,y)$ is well defined.
\begin{eqnarray}  \label{Green1}
\nabla \cdot \left(\frac{1}{ \varepsilon_s} \nabla G\right) 
+\omega^2 \mu_0 G &= \delta_y(x)&
\quad \textrm{in} \quad \mathbb R^2.
\end{eqnarray} 
with the  Sommerfeld radiation condition as $|x| \rightarrow +\infty$:
\bean \label{Green2}
\frac{\partial G}{\partial |x|} -i
\omega\sqrt{\varepsilon_0 \mu_0}G&=& O(\frac{1}{\sqrt{|x|}}),
\eean

A simple integration by parts in the system \eqref{u1}-\eqref{u2}
 yields

\bean\label{lippman}
H(x) = H_0(x) + \int_{B_\alpha} \left(\frac{1}{\varepsilon_m}- 
\frac{1}{\varepsilon_s(x)}\right) \nabla H(y) \nabla_yG(x,y) 
dy,
\eean
which leads to the following result. 
%%%%%%%%%%%%%%%%%%%%%%%%
\begin{proposition}
There exists a constant $C>0$, independent of $\alpha$
and $H_i$ such that 
\bea
\|H(x) -H_0(x)\|_{H^1(\Omega)}\leq C \alpha \|H_i\|_{H^1(\Omega)}. 
\eea
\end{proposition}
%%%%%%%%%%%%%%%%%%%%%
This proposition shows that if $\omega$ is real, the field
$H_0(x)$ is the  first term in the asymptotic expansion 
of $H(x)$ when $\alpha$ tends to zero. However, the constant 
$C$ in the proposition depend on $\varepsilon(x)$
and $\omega$ can be large. In fact, considering the results in
 proposition \eqref{asymp} and lemma 
\eqref{quasiresonance} if the
attenuation $\Gamma$ tends to zero the plasmonic resonances will 
approach the real axis and then the constant $C$ may blow up. In a
such situation one needs to take into account further terms in the
asymptotic expansion of $H(x)$ when $\alpha$ tends to zero in order
to improve the approximation. Here we
will derive  formally the first and  second terms  in the asymptotic expansion. 
In \cite{AK} an uniform asymptotic expansion of the 
magnetic field is derived using the method  of matched 
asymptotic expansions for $\alpha$ small enough.
Here we apply the same approach  to obtain a formal
asymptotic expansion of the electromagnetic 
fields.  We shall
represent the field $H(x)$  by two different expansions, an inner 
expansion for $x$ near $z^\star$, and an outer expansion for 
$x$ far away from $z^\star$. \\

The outer
expansion takes  the form
\bean\label{outer}
H(x) = H_0(x)+\alpha H_1(x) +\alpha^{2}H_2(x)+\cdots, 
\qquad \textrm{for} \quad |x-z^\star|>>O(\alpha),
\eean  
where  $H_1, H_2$ satisfy the following Helmholtz equation
\begin{eqnarray*} 
\nabla \cdot \left(\frac{1}{\varepsilon_s} \nabla H_i\right) +\omega^2 \mu_0 H_i &= 0&
\quad \textrm{in} \quad  |x-z^\star|>>O(\alpha),
\end{eqnarray*} 
with the  Sommerfeld radiation condition as $|x| \rightarrow +\infty$:
\bea
\frac{\partial H_i}{\partial |x|} -i\omega\sqrt{\varepsilon_0 \mu_0}H_i &=& O(\frac{1}{\sqrt{|x|}}).
\eea

Introducing 
the microscale variable $\xi = (x-z^\star)/\alpha$, then the inner expansion 
can be written as 
\bean \label{inner}
H(z^\star+\alpha \xi) = h_0(\xi)+\alpha h_1(\xi) +
\alpha^2\ln(\alpha) h_2(\xi)+\cdots, 
\qquad \textrm{for} \quad |\xi|=O(1),
\eean  
where the functions $h_0, h_1,  h_2$ satisfy the following  divergence form
equations:
\begin{eqnarray} \label{innerterms}
\nabla \cdot \left(\frac{1}{\tilde \varepsilon} \nabla h_0\right)  &= 0&
\quad \textrm{in} \quad  \mathbb R^2, \\
\nabla \cdot \left(\frac{1}{\tilde \varepsilon} \nabla h_1\right) + 
\nabla \cdot \left(\eta_1(\xi) \nabla h_0\right) 
&= 0&
\quad \textrm{in} \quad  \mathbb R^2,\\
\nabla \cdot \left(\frac{1}{\tilde \varepsilon} \nabla h_2\right)   
&= 0&
\quad \textrm{in} \quad  \mathbb R^2,
\end{eqnarray} 
where $\eta_1(\xi)$ and $\eta_2(\xi)$ are the coefficients of the inner expansion
of $\frac{1}{\varepsilon(z^\star+\alpha \xi) } $given by 

\bean\label{permitinner}
\frac{1}{ \varepsilon(z^\star+\alpha \xi)} = \frac{1}{\tilde \varepsilon(\xi)} 
+\eta_1(\xi)\alpha+\eta_2(\xi) \alpha^2+\cdots,
\eean
with
\bea
\eta_1(\xi) = \left\{ \ba{llcc}  \nabla(\frac{1}{\varepsilon_s})(z^\star) \xi  & \textrm{in}
&\mathbb R^2\setminus \overline B,\\
0& \textrm{in}
& B,
\ea 
\right.
\eea
and
\bea
\eta_2(\xi) = \left\{ \ba{llcc}  \nabla^2(\frac{1}{\varepsilon_s})(z^\star) \frac{\xi^2}{2}  & \textrm{in}
&\mathbb R^2\setminus \overline B,\\
0& \textrm{in}
& B,
\ea 
\right.
\eea
Obviously the inner and outer expansions are not valid everywhere and the 
systems of equations
satisfied by  the  functions $H_i$ and $h_i$ are not complete.  In order to
 determine these functions 
uniquely,  we need to equate the inner and the outer expansions in a some overlap 
domain within which the microscale variable $\xi$ is large and  $x-z^\star$ is small. 
In this domain the  matching conditions are:
\bea
H_0(y)+\alpha H_1(y)+\alpha^2 H_2(y) +\cdots \sim h_0(\xi)+\alpha
h_1(\xi)+ \alpha^2\ln(\alpha) h_2(\xi)
+\cdots
\eea

  A change of variables
in the Lippman-Schwinger integral representation formula 
\eqref{lippman} yields 
\bean \label{f1}
H(z^\star+\alpha \xi) =\\ H_0(z^\star+\alpha \xi) 
+\alpha \int_{B} \left(\frac{1}{\varepsilon_m}- 
\frac{1}{\varepsilon_s(z^\star+\alpha \xi^\prime)}\right) 
\partial_{\xi_k}\left(H(z^\star+\alpha \xi^\prime) \right)
\partial_{x_k}G(z^\star+\alpha \xi,z^\star+\alpha \xi^\prime)
d\xi^\prime.\nonumber
\eean
An asymptotic expansion  of the quantities above gives 
\bea
H_0(z^\star+\alpha \xi)  = H_0(z^\star) +\partial_{x_i}H_0(z^\star) \xi_i \alpha+
\partial_{x_ix_j}^2H_0(z^\star) \xi_i\xi_j \frac{\alpha^2}{2}+ o(\alpha^2),\\
\eea
and
\bea
\alpha\partial_{\xi_k}G(z^\star+\alpha \xi,z^\star+\alpha \xi) = 
\varepsilon_s(z^\star)\partial_{\xi_k}\Phi_0(\xi, \xi^\prime) +
\frac{1}{4\pi} \partial_{x_k}\varepsilon_s(z^\star) \alpha \ln(\alpha)
+\alpha \Phi_1(\xi, \xi^\prime) +o(\alpha),
\eea
where  $\Phi_0(\xi,\xi^\prime) = \frac{1}{2\pi}\ln(|\xi-\xi^\prime|)$ is the Green
function of the Laplacian in the whole space,   and  $
\Phi_1(\xi,\xi^\prime)$ is a weakly singular function (see 
Theorem~\ref{app1} in  Appendix).\\

  Inserting now  
the  inner expansion of $H$, and the above asymptotic expansion
into \eqref{f1} we obtain 
\bea
h_0(\xi) =  H_0(z^\star),\\
h_1(\xi)  =  \partial_{x_i}H_0(z^\star) \xi_i + \left(
\frac{\varepsilon_s(z^\star)}{\varepsilon_m} -1\right) 
\int_B \partial_{\xi_k}\Phi_0(\xi,\xi^\prime) \partial_{\xi_k} h_1(\xi^\prime) 
 d\xi^\prime,  
\eea
and
\bea
h_2(\xi) = 
  \left(
\frac{\varepsilon_s(z^\star)}{\varepsilon_m} -1\right) 
\int_B \partial_{\xi_k}\Phi_0(\xi,\xi^\prime)  \partial_{\xi_k} h_2(\xi^\prime) 
d\xi^\prime  +  
\frac{1}{4\pi} \left(\frac{1}{\varepsilon_m}-  \frac{1}{\varepsilon_s(z^\star)}\right) 
 \partial_{x_k}\varepsilon_s(z^\star)\int_B \partial_{\xi_k} h_1(\xi^\prime) 
d\xi^\prime.
\eea

Now we suppose that the functions $ h_0, h_1$ and 
$h_2$ are defined not just in the domain $B$, but everywhere in  $\mathbb
R^2$. Considering the asymptotic expansions obtained from 
the Lipmann-Schwinger equation and matching conditions, we obtain
\bean \label{h_0}
h_0(\xi) =  H_0(z^\star),
\eean 
\bean\label{h11}
\nabla \cdot \left(\frac{1}{\tilde \varepsilon} \nabla h_1(\xi) \right) 
&= 0& \textrm{in  }  \mathbb R^2,\\\label{h12}
\lim_{\xi\rightarrow +\infty}\left(h_1(\xi) -
\partial_{x_i}H_0(z^\star) \xi_i \right) &= 0,&  
\eean
and
\bean\label{h21}
\nabla \cdot \left(\frac{1}{\tilde \varepsilon} \nabla h_2 (\xi) \right) 
&= & 0 
\textrm{   in    }  \mathbb R^2,\\\label{h22}
\lim_{\xi\rightarrow +\infty}\left(h_2(\xi) - 
\frac{1}{4\pi} \left(\frac{1}{\varepsilon_m}-  \frac{1}{\varepsilon_s(z^\star)}\right) 
 \partial_{x_k}\varepsilon_s(z^\star)\int_B \partial_{\xi_k} h_1(\xi^\prime) 
d\xi^\prime\right) &= & 0. 
\eean

Using a variational approach  in the Hilbert space $W_0^{1, -1}(\mathbb 
R^2)$ one can prove that the systems \eqref{h12}- \eqref{h12} and 
\eqref{h22}- \eqref{h22}  have unique solutions.  Particularly, we
find that

\bean \label{exph2}
h_2(\xi) = \frac{1}{4\pi} \left(\frac{1}{\varepsilon_m}-\frac{1}{\varepsilon_s(z^\star)}\right) 
 \partial_{x_k}\varepsilon_s(z^\star)\int_B \partial_{\xi_k} h_1(\xi^\prime) 
d\xi^\prime,
\eean 
is indeed a constant function. \\

Now, we shall determine the  outer expansion functions $H_1$ and
$H_2$.  To do so we again  consider
the Lipmann-Schwinger equation
\bean \label{f2}
H(x) =\\ H_0(x) 
+ \alpha \int_{B} \left(\frac{1}{\varepsilon_m}- 
\frac{1}{\varepsilon_s(z^\star+\alpha \xi^\prime)}\right) 
\partial_{\xi_k}\left(H(z^\star+\alpha \xi^\prime) \right)
\partial_{x_k}\left(G(x,z^\star+\alpha \xi^\prime)\right) 
d\xi^\prime.\nonumber
\eean
Using the inner expansion of $H$ and the regularity of the Green 
function $G$ we obtain
\bean\label{outerHi}
H_1(x) = 0, \label{HH1} \\
 H_2(x) = \left(\frac{1}{\varepsilon_m} - 
\frac{1}{\varepsilon_s(z^\star)}\right) 
 \int_B \partial_{\xi_k}h_1(\xi^\prime) d\xi^\prime  
 \partial_{x_k}G(x,z^\star). \label{HH2} 
\eean

It is well known that the inner and outer expansions are not valid uniformly in  $x$ \cite{AK}.
 In order to obtain an asymptotic expansion of the fields as $\alpha$ tends to zero  that is valid 
 uniformly in  space variable, we merge to the two expansions together. Thus, adding the outer and 
inner expansions and subtracting out the common part, we formally  find
the following uniform expansions: for all  $x\in \Omega$: 
\bean
H(x) 
= H_0(x)+\alpha \mathcal 
H_1(\frac{x-z^\star}{\alpha})+\alpha^2\ln(\alpha) \mathcal 
H_2(\frac{x-z^\star}{\alpha}) + \alpha^2 H_2(x) +
O(\alpha^2\ln(\alpha)),
\eean
where 
\bea
 \mathcal H_1(\xi) =   h_1(\xi) -\xi_i
\partial_{x_i} H_0(z^\star) + 
\left(\frac{\varepsilon_s(z^\star)}{\varepsilon_m} -1 \right) 
\frac{1}{\pi} \int_{B} \partial_{\xi_i } h_1(\xi^\prime)  
d\xi^\prime \frac{\xi_i}{|\xi|^2}, \\
\mathcal H_2(\xi) =  \frac{1}{4\pi} \left(\frac{1}{\varepsilon_m}-\frac{1}{\varepsilon_s(z^\star)}\right) 
 \partial_{x_k}\varepsilon_s(z^\star)\int_B \partial_{\xi_k} h_1(\xi^\prime) 
d\xi^\prime.
\eea

Following the steps of the proof of Theorem 2.1 in \cite{AK} one can
obtain the following uniform asymptotic expansion. 
%%%%%%%%%%%%%%%%%%%%%%%%%%%%%%%%
 \begin{theorem}\label{asymfield} For $\delta \in(0, 1)$,
there exists a constant $C>0$, independent of $\alpha$
and $H_i$ such that 
\bea
\|H(x) -  H_0(x)-\alpha \mathcal H_1(\frac{x-z^\star}{\alpha})
 -\alpha^2\ln(\alpha) \mathcal 
H_2(\frac{x-z^\star}{\alpha}) -\alpha^2 H_2(x)\|_{H^1(\Omega)}\leq C
 \alpha^{2} \|H_i\|_{H^1(\Omega)}. 
\eea
\end{theorem}
The approximation can be improved by considering the inner 
expansion term of order $\alpha^2$ and computing the limit of 
$\Phi_1(\xi,\xi^\prime)$ as $\xi$ tends to $+\infty$.  Opposite
of the first impression, the
term $\alpha^2 H_2(x)$  on the right hand side is necessary
to cancel out the singularity of $\mathcal H_1(\xi)$ when $\xi$
tends to zero. Finally, if $\partial_{x_k}\varepsilon_s(z^\star) = 0$
one can recover the results of \cite{AK} by adding the 
 order $\alpha^2$ inner term.

%%%%%%%%%%%%%%%%%%%%%%%%%%%%%%%%%%%%%%%%%%%%% 
%%%%%%%%%%%%%%%%%%%%%%%%%%%%%%%%%%%%%%%
\subsection{The radial case}  \label{ssection4} Here we assume that
$\Omega$  and $B$ are the unit disc, and  $z^\star = 0$. We also
assume that the electric permittivity $\varepsilon$ is piecewise 
constant. \\

 Let $(r, \theta)$ be the polar coordinates in $\mathbb R^2$, $m$ be a
 fixed
integer larger than $1$, and  consider 
 $$H_i(r, \theta) = J_m(\frac{\omega}{c_0} r)e^{im\theta},$$ to be the
 magnetic incident field, where $J_m(\xi)$  is the Bessel 
function of the first kind of order $m$, and $c_0= 
\frac{1}{\sqrt{\varepsilon_0\mu_0}}$ is the speed of light in the 
free space.\\

Then, the total magnetic field takes the form $H(r,\theta) =
h_\alpha(r)e^{im\theta}$, with  
\bea
h(r) = \left\{\ba{lllccc}
\kappa_1 H_m(\frac{\omega}{c_0} r)+ 
J_m(\frac{\omega}{c_0} r) &\textrm{for } r\geq 1,\\
\kappa_2 H_m(\frac{\omega}{c_s} r)+ 
\kappa_3 J_m(\frac{\omega}{c_s} r)&\textrm{for } \alpha \leq r\leq 1,\\
\kappa_4 J_m(\frac{\omega}{c_m} r)& \textrm{for }  r\leq \alpha,
\ea
\right.
\eea 
where  $c_s= 
\frac{1}{\sqrt{\varepsilon_s\mu_0}}$, and  $c_m= 
\frac{1}{\sqrt{\varepsilon_m\mu_0}}$ are the speed of light 
in the dielectric coating and in the metallic nanoparticle
respectively.   $H_m(\xi)$  is the Hankel 
function of the first kind of order $m$. \\

The transmission
conditions for $r=1$ and $r=\alpha$ give the following system

\bea
\left(\ba{llllcccc}
H_m(\frac{\omega}{c_0}) &-H_m(\frac{\omega}{c_s})& 
-J_m(\frac{\omega}{c_s})&0\\
\frac{c_0}{c_s}
H_m^\prime(\frac{\omega}{c_0}) &-H_m^\prime(\frac{\omega}{c_s})& 
-J_m^\prime (\frac{\omega}{c_s})&0\\
0 & H_m(\frac{\omega}{c_s} \alpha)&J_m(\frac{\omega}{c_s} \alpha)&
-J_m(\frac{\omega}{c_m} \alpha)\\
0&\frac{c_s}{c_m}H_m^\prime
(\frac{\omega}{c_s} \alpha)&\frac{c_s}{c_m}
J_m^\prime(\frac{\omega}{c_s} \alpha) &
-J_m^\prime(\frac{\omega}{c_m} \alpha)\\
\ea
\right) \overrightarrow{\kappa} = 
\left(\ba{llll}-J_m(\frac{\omega}{c_0})\\
 -J_m^\prime(\frac{\omega}{c_0})
\\0 \\0
\ea \right).
\eea 

The plasmonic resonances, in this case, are exactly the zeros
of the determinant $d_\alpha(\omega)$, of the scattering matrix. 
An asymptotic expansion of the later when $\alpha $ tends to zero
gives 
\bea
d_\alpha(\omega) = \frac{d_0(\omega)}{\alpha} + o(\frac{1}{\alpha}), 
\eea

where 
\bea
d_0(\omega) := \left(-H_m(\frac{\omega}{c_0})J_m^\prime 
(\frac{\omega}{c_s}) + 
\frac{c_s}{c_0}
H_m^\prime(\frac{\omega}{c_0})J_m(\frac{\omega}{c_s})\right) 
\frac{c_s^m}{\pi\omega}(c_m^2+ c_s^2)\frac{1}{c_m^{m+1}}.
\eea
Hence a limiting value $\omega(0)$ of a sequence of plasmonic 
resonances has to be finite and satisfies the dispersion
equation $d_0(\omega(0)) = 0$. We remark that the complex roots
of  the function 
\bea 
-H_m(\frac{\omega}{c_0})J_m^\prime 
(\frac{\omega}{c_s}) + 
\frac{c_0}{c_s}
H_m^\prime(\frac{\omega}{c_0})J_m(\frac{\omega}{c_s}),
\eea
are exactly  the scattering resonances of the domain $\Omega$ in absence 
of the nanoparticle. If we drop the assumption that $\omega$
is small, and if the material that fills the nanoparticle is 
non-dispersif,  we obtain the well known  
convergence of the scattering resonances to the non perturbed
ones (see for instance \cite{Ammari2, AK}). \\

A
careful analysis of the zeros of $d_0(\omega)$ in the quasi-static 
regime leads to $\varepsilon_m(\omega(0)) = -\varepsilon_s$
or $\varepsilon_m(\omega(0))= 0$, which 
correspond  exactly to the plasmonic values of the circular shape nanoparticle 
$\beta_\infty^\pm= \frac{1}{2}$ and $\beta_1^-=0$ (see for instance 
\eqref{dispersion}). \\

In the case where
$m$ is equal to one the determinant $d_\alpha(\omega)$ has  the following asymptotic 
expansion 
$d_\alpha(\omega)= d_0(\omega)\ln(\alpha)+o(\ln(\alpha))$ as $\alpha$ 
tends to zero.  Using Rouch\'e theorem one can determine the complete asymptotic
expansion of the plasmonic resonances in the case of a circular shape.

%%%%%%%%%%%%%%%%%%%%%%%%%%%%%%%%%%%%%%%
\section{Photoacoustic effect}
\label{section2}
In this section  we consider a metallic  nanoparticle  in a liquid medium and we want to describe the photoacoustic generation created by the electromagnetic heating of the nanoparticle. We derive the 
model equations that describe the coupling between the temperature rise in the medium and the 
acoustic wave generation. 

\subsection{Acoustic sources}

We write the fundamental equations of acoustics as explained in \cite{bossy}, i.e the equation of continuity, the Euler equation and the continuity equation for heat flow.

\begin{equation}
\label{ac1}
\frac{\partial \rho}{\partial t} = -\rho_0 \text{div}(v),
\end{equation}

\begin{equation}
\label{ac2}
\rho_0\frac{\partial v}{\partial t} = -\nabla p,
\end{equation}

\begin{equation}
\label{entr}
\rho_0 T\frac{\partial s}{\partial t} = \text{div}(\kappa \nabla T) + P_v,
\end{equation}

where $\rho$ is the mass density, $p(r,t)$ is the acoustic pressure, $v(r,t)$ is the acoustic displacement velocity, $s(r,t)$ is the specific entropy, $T(r,t)$ is the temperature and  $P_v$ is the heat source. The change of density is assumed small ($\frac{\rho-\rho_0}{\rho_0}\ll 1$).  The thermal conduction $\kappa$  is given by 
\bea
\kappa(x) \;=\; \left\{\ba{llcc} 
\kappa_s(x) \quad &\textrm{for   }  x\in \Omega \setminus 
\overline{B_\alpha},\\
\kappa_0 \quad &\textrm{for   } 
x\in B_\alpha,
 \ea \right.
\eea
where $\kappa_s(x)>0$ is the thermal conduction of the liquid and $\kappa_0>0$ is the thermal conduction
of the metal that fills the nanoparticle, and verifies  $\kappa_0>>\kappa_s$. \\

 We can write the two equations of state giving the change of density $\delta \rho$ and the change of entropy $\delta s$ in terms of $\delta p$ and $\delta T$ \cite{morse}.

\begin{equation}
\label{drho}
\delta \rho = \frac{\gamma}{c_s^2} \delta p - \rho_0 \beta \delta T,
\end{equation}

\begin{equation}
\label{ds}
\delta s = \frac{c_p}{T}(\delta T - \frac{\gamma-1}{\rho_0\beta c_s^2}\delta p),
\end{equation}

where $c_p = T\left(\frac{\partial s}{\partial T}\right)_p$ is the specific heat capacity at constant pressure, $c_v = T\left(\frac{\partial s}{\partial T}\right)_{\rho}$ is the specific heat capacity at constant volume, $\gamma=\frac{c_p}{c_v}$, $\beta=-\frac{1}{\rho}\left(\frac{\partial \rho}{\partial T}\right)_p$ is the thermal expansion coefficient, and $c_s$ is the isentropic sound velocity.\\

We deduce from equation \eqref{drho} and \eqref{ds} the two following equations:

\begin{equation}
\frac{\partial \rho}{\partial t} = \frac{\gamma}{c_s^2} \frac{\partial p}{\partial t} - \rho_0 \beta \frac{\partial T}{\partial t},
\end{equation}

\begin{equation}
\label{dsdt}
\frac{\partial s}{\partial t} = \frac{c_p}{T}(\frac{\partial T}{\partial t} - \frac{\gamma-1}{\rho_0\beta c_s^2}\frac{\partial p}{\partial t}).
\end{equation}

We can make the assumption for liquids that $\gamma = 1$. With this assumption and combining equations \eqref{entr} and \eqref{dsdt}, we obtain the following equation for the temperature field $T$:

\begin{equation}
\rho_0 c_p\frac{\partial T}{\partial t} = \text{div}(\kappa \nabla T) + P_v.
\end{equation}

We now use equations \eqref{ac1} and \eqref{ac2} to get $\frac{\partial ^2 \rho}{\partial t^2} - \Delta p = 0$. We can transform this equation thanks to equation \eqref{drho} and we obtain:
\begin{equation}
\frac{\gamma}{c_s^2}\frac{\partial ^2 p}{\partial t^2} - \Delta p = \rho_0 \frac{\partial}{\partial t}\left( \beta \frac{\partial T}{\partial t}\right).
\end{equation}

With the assumption that $\gamma = 1$ and that $\beta = \beta_0$, we finally have the following system of coupled equations for the generation of photoacoustic waves in a liquid medium:
\begin{equation}
\rho_0 c_p\frac{\partial T}{\partial t} = \text{div}(\kappa \nabla T) + P_v,
\end{equation}
\begin{equation}
\frac{1}{c_s^2}\frac{\partial ^2 p}{\partial t^2} - \Delta p = \rho_0 \beta_0 \frac{\partial^2 T}{\partial t}.
\end{equation}

\subsection{Electromagnetic sources}

The source term $P_v$ in equation \eqref{entr} is the energy produced by 
electromagnetic heating. It can be written as follows \cite{pearce}:

\begin{equation}
P_v = Q_{gen} + Q_{met},
\end{equation}

where $Q_{gen}$ is the volumetric power density of the electromagnetic source, and $Q_{met}$ is the metabolic heat generated by biological tissues. We consider here that $Q_{met}=0$.

The electromagnetic coefficients of the medium are the complex electric permittivity $\varepsilon_s$, the magnetic permeability $\mu_0$. Since the electromagnetic wave is time pulsed and because of the  difference of 
time scales between the acoustic and electromagnetic waves, 
the volumetric power density is described by  the time averaging of the real part of the divergence of the Poynting vector $\textbf{S}=\textbf{E}\times \overline{\textbf H}$ times the Dirac function
at zero.  On the other hand  the  divergence of $\textbf{S}$ is given by

\begin{equation}
-\nabla\cdot \textbf{S} = i\omega\overline \varepsilon|\textbf{E}|^2 + i\omega \overline 
\mu_0 |\textbf{H}|^2.
\end{equation}
By the  taking the real part and time averaging of the divergence of the  Poynting vector
 we finally have

\bean
Q_{gen} = \omega \Im(\varepsilon) \langle|\textbf{E}|^2\rangle \delta_0(t)
= \omega \Im(\varepsilon)  |E|^2 \delta_0(t),
\eean

where  the time averaging  is defined by 
 $ \langle f \rangle := \lim_{\tau \rightarrow +\infty} \int_0^\tau f(t) dt$, and $\delta_0$ is the Dirac 
 function at $0$.  \\

We can finally write the following system of coupled equations that describes the photoacoustic generation by the electromagnetic heating of a metallic nanoparticle

\bean
\rho_0 c_p\frac{\partial T}{\partial t}& =& \text{div}(\kappa \nabla T) + \omega\Im(\varepsilon)|E|^2 \delta_0(t),\\
\frac{1}{c_s^2}\frac{\partial ^2 p}{\partial t^2} - \Delta p & =& \rho_0 \beta_0 \frac{\partial^2 T}{\partial t}.
\label{acous}
\eean
with  the initial conditions at  $t=0$:
\bean
T= p= \frac{\partial p}{\partial t} = 0.  
\eean

Following the same analysis as in   \cite{AKV} one cane show that   the temperature  $T$ 
approaches  $T_0$ as $\alpha$ tends to zero,  where $T_0$ is the solution to 
\bea
\rho_0 c_p\frac{\partial T_0}{\partial t} =\text{div}(\kappa_s \nabla T_0) + 
\omega\Im(\varepsilon)|E|^2 \delta_0(t),
\eea
with initial  boundary condition $T_0= 0$ at $t=0$, and $\lim_{|x|\to +\infty}T_0(x) = 0$. 
 Here we did not consider the first and second
terms in the small volume asymptotic expansion because of the absence the limiting problems
are well posed compared with the ones in the asymptotic expansion of the EM fields.\\

Since the conductivity  $\kappa_s$  of the biological is very small compared to the
other quantities we neglect it and find  the following equation for the temperature
\bea
\rho_0 c_p\frac{\partial T_0}{\partial t} = \omega\Im(\varepsilon)|E|^2 \delta_0(t),
\eea
which combined with the acoustic waves \eqref{acous},  provides  at the end
the following model for the photo acoustic effect by a metallic  nanoparticle:
\bean \label{mainequation3}\left\{\ba{llcc}
\frac{1}{c_s^2}\frac{\partial ^2 p}{\partial t^2}(x,t) - \Delta p(x,t) =  0 &\textrm{in} &
\mathbb R^2 \times \mathbb R_+,\\
p(x,0) = \frac{\omega\beta_0}{c_p}\Im(\varepsilon)(x)|E(x)|^2&\textrm{in} &
\mathbb R^2,\\
\frac{\partial p}{\partial t}(x,0) =0 &\textrm{in} &
\mathbb R^2.\\
\ea
\right.
\eean
 The 
system above  \eqref{mainequation3} coupled with the Helmholtz equation 
\eqref{mainequation1}-\eqref{mainequation2}   represents
the forward problem. Next, we study the photoacoustic  inverse problem.
%%%%%%%%%%%%%%%%%%%%%%%%%%%%%%%%%%%%%
\section{The photoacoustic inverse problem}
%%%%%%%%%%%%%%%%%%%%%%%%%%%%%%%%%%%%%%
In this section we study the inverse problem of the reconstruction 
of the electric permittivity $\varepsilon$ from the measurements of the
acoustic waves $p(x,t),\,  (x,t)\in \partial \Omega\times (0, \tau_p)$, generated
by the photoacoustic effect  from the heating of the small metallic nanoparticle $B_\alpha$ 
in the presence of  electromagnetic fields at a frequency close to a plasmonic resonance.  Here 
  The constant $\tau_p>0$ is  
the period of time where the measurements are taken, that will be specified later.
We have two inversions: the  acoustic inversion where we assume that the speed of the wave 
is  a known constant  $c_s$ and reconstruct the initial pressure   
$\Im(\varepsilon(x))|E(x)|^2, \, x\in \Omega$
from the knowledge of $p(x,t),\,  (x,t)\in \partial \Omega\times (0, \tau_p)$; the second 
step is to recover the electric permittivity $\varepsilon$ from the internal data 
$\Im(\varepsilon(x))|E(x)|^2, \, x\in \Omega$.
%%%%%%%%%%%%%%%%%%%%%%%%%%%%%%%%%%%%%%%%%%
%%%%%%%%%%%%%%%%%%%%%%%%%%%%%%%%%%%%%%%%%%%

\subsection{Acoustic inversion}
Recall that $\Im(\varepsilon)(x)$ is  a compactly supported function in $\Omega$, and that
we have assumed that the acoustic wave speed  in the tissue takes a constant 
value $c_p$ that corresponds to the  isentropic  acoustic speed  in
the water, that is 1400 m/s. These two assumptions allow us to use  
well know results from  control theory  to
derive a stability estimate for the acoustic inversion. 
 The following result is based on the multiplier method 
and can be found in \cite{Ho, Lions2}.

%%%%%%%%%%%%%%%%%%%%
\begin{theorem} \label{resul}
Let $\tau_p > \tau_\Omega$ where  $\tau_p= \sup_{x,y\in \Omega }|x-y|$. Then,
there exists a constant $C= C(\Omega)>0$ such  that 
\bea 
\frac{\omega\beta_0}{c_p}\|\Im(\varepsilon(x))|E(x)|^2\|_{L^2(\Omega)}
\leq C \| \frac{\partial p}{\partial t} \|_{L^2(\partial \Omega\times
(0,\tau_p))} + \| \nabla p\|_{L^2(\partial \Omega\times (0,\tau_p))}
\eea
\end{theorem}
%%%%%%%%%%%%%%%%%%

We refer the readers to the survey \cite{KuKu-EJAM08} on related
reconstruction methods and different approaches based on  integral equations
for constant acoustic speed. 
The stability result shows  that the reconstruction of the electromagnetic
energy responsible  for the generation of  the acoustic signal by heating 
the nanoparticle, from boundary measurements of the acoustic waves  is stable if  the
observation time  $\tau_p$ is large enough. This result can be
extended to a non constant acoustic speed as well as measurements of the
acoustic waves on  a small part of the boundary
\cite{AmBoJuKa-SIAM10, PU, Kuchment}. 
In this paper for the sake of simplicity we do not handle such general cases.  \\        

We further  assume that  the constants  $\beta_0$ and $c_p$ are given. 
Let $\mathcal O_M$  denotes    the  ball centered at $0$ and of  radius $M>0$ 
in $H^2(B_R(z^\star))$, where $R>0$ is large enough such that $\overline 
\Omega \subset B_R(z^\star)$.

%%%%%%%%%%%%%%%%%%%%%%%%%%%%%%%
\begin{corollary} \label{main2bis}Assume that $\varepsilon \in B_M(0)$, and let 
$\tau_p > \tau_\Omega$. Then,  there exists a constant 
$C= C(\omega, M, \beta_0, c_p)>0$  such that the following estimate
\bean \label{stabacous}
\|\Im(\varepsilon)|\nabla H|^2\|_{C^0(\overline\Omega)}
\leq C \left(\left\| \frac{\partial p}{\partial t} \right\|_{L^2(\partial \Omega\times
(0,\tau_p))} + \| \nabla p\|_{L^2(\partial \Omega\times
 (0,\tau_p))}\right)^{\frac{1}{4}},
\eean
holds. 
\end{corollary}
%%%%%%%%%%%%%%%%%%%%%%%%%%%%
\proof 
A simple calculation yields $|E(x)|^2 = |\nabla H(x)|^2$ over
$\Omega$. Using the interpolation
between  Sobolev spaces  \cite{Lions1},  we estimate  $\Im(\varepsilon)|\nabla H|^2$  in  $H^{\frac{3}{2}}(\Omega)$ in terms of its norms in 
$L^2(\Omega)$ and $H^2(\Omega)$ respectively. Thus we deduce $\eqref{stabacous}$
from Elliptic regularity  of the system \eqref{mainequation1} and  the  estimate  in theorem
\eqref{resul}.

\endproof

%%%%%%%%%%%%%%%%%%%%%%%%%%%%%%%%%%%%%%%%%%
%%%%%%%%%%%%%%%%%%%%%%%%%%%%%%%%%%%%%%%%%%%
\subsection{Optical inversion}
In this part of the paper  we assume that the  internal electromagnetic energy
  $$\Im(\varepsilon(x))|\nabla H(x)|^2,$$ for $
\, x\in \Omega$ is  recovered, and we study the inverse problem of 
determining   $\varepsilon(x)$ over  $ \Omega$ using  the small
volume asymptotic
expansion of the EM fields in the previous section. In fact in
applications
we only need to recover the imaginary part of the electric
permittivity which is related to the absorption of the EM fields
and the generation of the photoacoustic wave.  \\

Recall that the absorption of EM energy by only the biological tissue 
is negligible inside $\Omega$. In practice 
the photoacoustic signal 
generated by such absorption is weak inside $\Omega$ and can 
not be used to image the tissue itself.  \\
%Thus we assume that
%$\Im(\varepsilon(x))|\nabla H_0(x)|^2\approx 0$ over $\Omega$.\\

From section 2 we deduce  the inner and outer asymptotic expansions of
the magnetic field $|\nabla H(x)|^2$. Our strategy here is to first
analyze the  information  about the medium and
the nanoparticle contained in   the outer asymptotic  expansion. This 
problem  is a classical boundary/internal   inverse problem,
and has some known limitations.  Then  we complete  the recovery
of the optical properties of the medium using  information retrieved 
from the inner expansion of the magnetic field and the  apriori 
information about the shape of the nanoparticle.  
%%%%%%%%%%%%%%%%%%%%%%%%%%%%%%
\subsubsection{Inversion using the outer expansion} \label{outerexpansion}
%%%%%%%%%%%%%%%%%%%%%%%%%%%%%%%%

Recall the outer asymptotic expansion \eqref{outer}-
\eqref{outerHi} of the magnetic field: 
\bea
H(x) = H_0(x) +\alpha^2H_2(x) +o(\alpha^2) \qquad 
\textrm{for   } x\in \partial \Omega,
\eea
where $H_0(x)$ is the solution to the system \eqref{u1}-
\eqref{u2}, and $H_2(x)$  is given
by
\bea 
H_2(x) = \left(\frac{1}{\varepsilon_m(\omega)} - 
\frac{1}{\varepsilon_s(z^\star)}\right) 
 \int_B \partial_{\xi_k}h_1(\xi^\prime) d\xi^\prime  
 \partial_{x_k}G(x,z^\star), 
\eea
with $h_1(\xi)$ is the unique solution to the system 
\eqref{h11}-\eqref{h12}. \\

In fact the asymptotic expansion above
is valid in a neighboring region of the boundary $\partial \Omega$,
but since the internal data  is of the  form
$\Im(\varepsilon(x))|\nabla H(x)|^2$, where $\Im(\varepsilon)$
is compactly supported in $\Omega$,  we can only retrieve information
about the magnetic field on the boundary $\partial \Omega$. Note that since 
$\varepsilon_0 $  is given  one can retrieve the  the Cauchy 
data of the magnetic field on $\partial \Omega$ form the knowledge 
of its trace on the same set. \\

The function $H_2(x)$ can  be rewritten in
terms of the first order polarization tensor  
$M(\frac{\varepsilon_m(\omega)}{\varepsilon_s(z^\star)})  = (M_{kl})_{1 \leq
  k, l\leq 2}$, as follows (see for instance \cite{AmmariKang} 
and references therein)
\bea
H_2(x) =  \left(\frac{1}{\varepsilon_m(\omega)} - 
\frac{1}{\varepsilon_s(z^\star)}\right)  
\nabla G(x,z^\star)\cdot M\nabla H_0(z^\star),
\eea
where 
\bean \label{pt}
M_{kl} = 
\int_B \partial_{\xi_k} \phi_l(\xi^\prime) d\xi^\prime, 
\eean
and $\phi_l(\xi), l=1,2$ are the unique solutions to
the system
\bean\label{phil1}
\nabla \cdot \left(\frac{1}{\tilde \varepsilon} \nabla \phi_l(\xi) \right) 
&= 0& \textrm{in  }  \mathbb R^2,\\\label{phil2}
\lim_{\xi\rightarrow +\infty}\left(\phi_l(\xi) -\xi_l \right) &= 0.&
\eean

On the other hand   $\phi_l(\xi), l=1,2$  can  be rewritten
as follows
\bean \label{phileq}
\phi_l(\xi) =  \xi_l - \left(\frac{\varepsilon_m(\omega)}{\varepsilon_s(z^\star)
-\varepsilon_m(\omega)}I+\mathcal T_0\right)^{-1} \hat{\xi}_l (\xi),
\eean
where $\hat{\xi}_l(\xi)\in  W^{1,-1}_0 (\mathbb R^2)$ is 
the orthogonal projection of $\xi_l\chi_B(\xi)$ 
onto $W^{1,-1}_0 (\mathbb R^2)$, which can be defined as  the unique solution to 
the system 
\bea
\int_{\mathbb R^2 }\nabla  \hat{\xi}_l \nabla v d\xi  =    \int_{B }
\nabla \xi \nabla v d \xi   \qquad \textrm{for  all    } v\in  
 W^{1,-1}_0 (\mathbb R^2).
\eea

Regarding the integral equation \eqref{phileq}, we observe
than when $\omega $ tends to a  plasmonic resonance $\omega_j(\alpha)$ 
the functions $\phi_l(\xi)$, and consequently the polarization tensor
$M$ will most likely blows up.  Since  in applications $\omega$ 
is real, and the plasmonic resonances of the  nanoparticle embedded 
in the medium  approaches the quasi-static resonances 
$\omega_j(0)$ when $\alpha$ tends to zeo (proposition \eqref{asymp}, 
we expect that the coefficient $M$ become large in the case where   
$\omega$  coincides with $\Re(\omega_j(0))$, and $\Gamma <<1$. \\

Many works have considered the  localization of small 
inhomogeneities in a known background 
medium, and most of the  proposed methods are  based
on an appropriate averaging of the asymptotic expansion
 by using particular  background solutions  as 
weights~\cite{AmmariKang, AMV}. In other words,  the position 
$z^\star$ of the nanoparticle 
can be uniquely determined from the  outer expansion of $H(x)$, that 
is    $H_0(x)+\alpha^2 H_2(x),\; x\in \partial \Omega$,
if  the electric permittivity of the background medium $\varepsilon_s(x)$ is 
known everywhere. But  this  is not the case in our problem, since our
objective is to determine $\varepsilon_s(x)$, while
$\varepsilon_m(\omega)$
is known (which is the complete opposite of the setting
where small inhomogeneities are imaged).  Here to overcome these 
difficulties we may propose the use of multifrequency measurements $H_2(x), 
\omega\in(\underline \omega, \overline \omega)$  to localize $z^\star$ \cite{ACZ, Garnier}, 
where $\underline{\omega}, \overline{\omega}$  are two strictly positive
constants satisfying  $\underline{\omega}<<\overline{\omega}$. We 
will study this  specific  inverse problem in future works. From now
on
we assume that the position $z^\star$ of the nanoparticle is known.      \\

Note that in general  if   $\varepsilon_s(x)$ 
is known, it is still  not possible to recover simultaneously the shape of the
nanoparticle  $\partial B$ and the contrast
$\frac{\varepsilon_s(z^\star)}{\varepsilon_m(\omega)}$ from only the
measurement of the outer expansion $H_0(x)+\alpha^2 H_2(x),\;
x\in \partial \Omega$. Meanwhile  in our setting the shape of the 
nanoparticle  is assumed to be known. For example, if we consider 
the circular shape, that is $B$ is  the unit disc,
$\hat{\xi}_l(\xi),\, l=1,2, $ and hence 
$\phi_l(\xi), \, l=1,2, $ can be determined explicitly
\bean 
 \hat{\xi}_l(\xi)= \left\{ \ba{llcc}
 \frac{\xi_l}{2} &\textrm{for}\; \xi \in B,\\ 
 \frac{\xi_l}{2|\xi|^2} &\textrm{for}\; \xi \in \mathbb R^2\setminus \overline B,
 \ea  \right. \\\label{philxi}
\phi_l(\xi) =  \left\{ \ba{llcc}
  \frac{2\varepsilon_m(\omega)}{\varepsilon_s(z^\star)
+\varepsilon_m(\omega)} \xi_l &\textrm{for}\; \xi \in B,\\ 
  \xi_l-\frac{\varepsilon_s(z^\star)-\varepsilon_m(\omega)}{\varepsilon_s(z^\star)
+\varepsilon_m(\omega)} \frac{\xi_l}{|\xi|^2} &\textrm{for}\; \xi \in 
\mathbb R^2\setminus \overline B,
 \ea  \right.
\eean
which implies that   
the polarization tensor can be simplified  into
\bea
M_{kl} = \frac{2\varepsilon_m(\omega)}{\varepsilon_s(z^\star)
+\varepsilon_m(\omega)} |B |\delta_{kl},
\eea
where $\delta_{kl}$ is Kronecker  symbol.  Assuming that 
$H_0(x),\; x\in \partial \Omega$ 
is given, we deduce from the outer expansion the following
approximation 
\cite{AmmariKang, AMV}:
\bean
\frac{1}{\varepsilon_0}
 \int_{\partial \Omega} \left(H\frac{\partial H_0}{\partial \nu_\Omega} 
- \frac{\partial H}{\partial \nu_\Omega}H_0 \right) ds(x) \\
=\alpha^2 \left(\frac{1}{\varepsilon_m(\omega)} - 
\frac{1}{\varepsilon_s(z^\star)}\right)  
\nabla  H_0(z^\star) \cdot M\nabla H_0(z^\star) +o(\alpha^2)\nonumber \\
= 2|B| 
\frac{\varepsilon_s(z^\star)-\varepsilon_m(\omega)}{
\varepsilon_m(\omega)+ \varepsilon_s(z^\star)}  
\frac{1}{\varepsilon_s(z^\star)}\left|\nabla H_0(z^\star)\right|^2 
\alpha^2   +o(\alpha^2),
\label{cs}
\eean 

To ensure that the first term of the asymptotic expansion
does not vanish, and to guarantee the success of the identification 
procedure it becomes necessary to assume  the following non-degeneracy 
condition
$$ \left|\nabla H_0(z^\star)\right|^2 \not=0.$$ 
For a circular shape nanoparticle we can immediately see  from the 
explicit  expression
of the first term in the asymptotic expansion  that when $\omega$ is 
close to a  plasmonic resonance, that is
$\varepsilon_m(\omega) = -\varepsilon_s(z^\star)$, the polarization 
tensor constant blows up. In the  next paragraph we  investigate the 
inner expansion  of the magnetic field which represents our  
photoacoustic data,  in order  to derive the contrast
$\frac{\varepsilon_s(z^\star)}{\varepsilon_m(\omega)}$.

 %%%%%%%%%%%%%%%%%%%%%%%%%%%%%%%%%%%
\subsubsection{Inversion using the inner expansion} \label{innerexpansion}
%%%%%%%%%%%%%%%%%%%%%%%%%%%%%%%%%%%
We further assume that the position $z^\star$,  the size $\alpha$
and the shape $\partial B$ of
the nanoparticle are known. Recall the inner expansion \eqref{inner}:

\bea
H(z^\star+\alpha \xi) 
= H_0(z^\star)+\alpha h_1(\xi) +
\alpha^2\ln(\alpha) h_2(\xi) +O(\alpha^2) 
\qquad \textrm{for} \quad |\xi|=O(1).
\eea  
where $h_1(\xi)$  is the unique solution to the system
 \eqref{h11}-\eqref{h12}, that is 
\bea
\nabla \cdot \left(\frac{1}{\tilde \varepsilon} \nabla h_1(\xi) \right) 
&= 0& \textrm{in  }  \mathbb R^2,\\
\lim_{\xi\rightarrow +\infty}\left(h_1(\xi) -
\partial_{x_i}H_0(z^\star) \xi_i \right) &= 0,&  
\eea
and  $h_2(\xi)$ is a constant fucntion given by
\bea
h_2(\xi) = \frac{1}{4\pi} \left(\frac{1}{\varepsilon_m}-
\frac{1}{\varepsilon_s(z^\star)}\right) 
 \partial_{x_k}\varepsilon_s(z^\star)\int_B \partial_{\xi_k} h_1(\xi^\prime) 
d\xi^\prime.
\eea

Using the functions $\phi_l, l=1,2  $ solutions to the
system \eqref{phil1}-\eqref{phil2}, we can rewrite $h_1(\xi)$
as
\bean \label{h1rewritten}
h_1(\xi) = \phi_k(\xi) \partial_{x_k} H_0(z^\star). 
\eean

Recall  that the   acoustic inversion  provides the 
 internal function $ \Psi(x)=\Im(\varepsilon(x))|\nabla H(x)|^2,\; x\in \Omega$. 
Combining   \eqref{inner} and \eqref{permitinner}, we obtain
 the following inner expansion  
\bean
\Psi(z^\star+\alpha \xi)&=& 
\Im(\varepsilon(z^\star+\alpha \xi))|\nabla H(z^\star+\alpha \xi)|^2
\nonumber\\
&=& \Im(\tilde \varepsilon_0(\xi))|\nabla_\xi h_1(\xi)|^2 +
 O(\alpha^2),\qquad \textrm{for} \quad |\xi|=O(1).
 \label{datainner}
\eean
We further assume that $B$ is the unit disc.  Our objective is to
recover $\varepsilon_s(z^\star)$ from the knowledge of 
$ \Im(\tilde \varepsilon_0(\xi))|\nabla_\xi h_1(\xi)|^2$ for $\xi \in 2B$, where 
$2B$ is the disc of center zero and radius $2$. \\
 
Combining 
\eqref{h1rewritten} and \eqref{philxi}, we find  
\bea
h_1(\xi) =  \left\{ \ba{llcc}
  \left(1-\kappa \right) \xi\cdot \nabla H_0(z^\star) &\textrm{for}\; \xi \in B,\\ 
   \left(1-\frac{\kappa}{|\xi|^2} \right) \xi \cdot \nabla H_0(z^\star)&\textrm{for}\; \xi \in 
2B\setminus \overline B,
 \ea  \right.
\eea
where 
\bea
\kappa := \frac{\varepsilon_s(z^\star)-\varepsilon_m(\omega)}{\varepsilon_s(z^\star)
+\varepsilon_m(\omega)}.
\eea
Hence
\bea
\Psi(z^\star+\alpha \xi) +o(\alpha) = \left\{ \ba{llcc}
\Im(\varepsilon_m(\omega))|1-\kappa|^2 |\nabla H_0(z^\star)|^2 &\textrm{for}\; \xi \in B,\\ 
 \Im(\varepsilon_s(z^\star))\left|\nabla_\xi\left( \left(1-
 \frac{\kappa}{|\xi|^2} \right) \frac{\xi}{|\xi|^2} \cdot \nabla H_0(z^\star)
\right)\right|^2&\textrm{for}\; \xi \in 
2B\setminus \overline B.
 \ea  \right. 
\eea

A forward  calculation yields 
\bea
\Psi(z^\star+\alpha \xi)  & = &
\Im(\varepsilon_s(z^\star))\left|
\left(1-\frac{\kappa}{|\xi|^2} \right) \nabla H_0(z^\star) +
2\kappa \frac{\xi}{|\xi|^2}\cdot \nabla H_0(z^\star)\frac{\xi}{|\xi|^2}
\right|^2+O(\alpha),
\eea
for $  \xi \in 
2B\setminus \overline B$.\\

Now taking the ratio between $\Psi|^+_{\partial B_\alpha}$ and
$\displaystyle \Psi(z^\star)
= \fint_{B_\alpha}\Psi(x) dx$, we obtain
\bean
\frac{\Psi(z^\star+\alpha \xi)|_+}{\Psi(z^\star)} &= &
\frac{\Im(\varepsilon_s(z^\star))}{\Im(\varepsilon_m(\omega))}
\left( \left|\frac{1+\kappa}{1-\kappa}\right|^2 
\left|\frac{\nabla H_0(z^\star)}{| \nabla H_0(z^\star)|}\cdot \xi \right|^2
+ \left|\frac{\nabla H_0(z^\star)}{| \nabla H_0(z^\star)|}
\cdot \xi^\perp \right|^2\right) +O(\alpha),\nonumber\\
&=& \Psi_0(\xi) +O(\alpha),\label{psi0}
\eean
for $  \xi \in 
\partial B=\left\{  \xi^\prime
\in \mathbb R^2; \;\; |\xi^\prime| =1\right\} $, where $\xi^\perp$ is a $\frac{\pi}{2}$ 
counterclockwise  rotation of $\xi$.\\ 

Now, assuming that $|\Re(\varepsilon_m(\omega))|>|\Re(\varepsilon_s(z^\star)) |$, 
we have
\bea
 \left|\frac{1+\kappa}{1-\kappa}\right|>1,
\eea
 and thus the function $\Psi_0(\xi)$ takes its maximum and minimum on 
$\partial B$
 at $\xi= \pm \frac{\nabla H_0(z^\star)}{| \nabla H_0(z^\star)|}$ and  
 $\xi= \pm \frac{\nabla H_0(z^\star)^\perp}{| \nabla H_0(z^\star)|}$  respectively. \\

Consequently
\bean \label{i21}
\frac{\Im(\varepsilon_s(z^\star))}{\Im(\varepsilon_m(\omega))} 
&=& \frac{\Psi(z^\star+\alpha \frac{\nabla H_0(z^\star)^\perp}
{| \nabla H_0(z^\star)|})|_+}{\Psi(z^\star)}  +O(\alpha), \\&=& 
\min_{\xi\in \partial B} 
\frac{\Psi(z^\star+\alpha \xi)|_+}{\Psi(z^\star)}  +O(\alpha),  \label{i22}
\eean
and
\bean \label{j21}
\frac{\Im(\varepsilon_s(z^\star))}{\Im(\varepsilon_m(\omega))} 
\left|\frac{1+\kappa}{1-\kappa}\right|^2  
&=& \frac{\Psi(z^\star+\alpha \frac{\nabla H_0(z^\star)}
{| \nabla H_0(z^\star)|})|_+}{\Psi(z^\star)}  +O(\alpha), \\&=& 
\max_{\xi\in \partial B} 
\frac{\Psi(z^\star+\alpha \xi)|_+}{\Psi(z^\star)}  +O(\alpha),  \label{j22}
\eean

Since $ \varepsilon_m(\omega)$ is given, we can retrieve
$\Im(\varepsilon_s(z^\star))$ from   equality \eqref{i21}, and   
then $\Re(\varepsilon_s(z^\star))$ from  equality \eqref{i22}. Now, we are
able to prove  the main theorem stability estimate.   
%%%%%%%%%%%%%%%%%%%%%%%%%%%%%%%%%%%%%%%%%%%%%%%%%%%%%%%%%%%%%%%%%%%%%%
\subsection{Proof of the main theorem~\eqref{main}} 

We deduce from equalities  \eqref{i21}-\eqref{i22} the following
estimates.

\begin{theorem} \label{main3}
Under the  same assumptions as in theroem~\eqref{main}, there exists
a constant $C>0$ that does not depend on $\alpha$, such that
\bea
|\Im(\varepsilon_{s,a}(z^\star)) - 
\Im(\varepsilon_{s,a}(z^\star))| \leq 
C\|\Psi_a -\Psi_b\|_{L^\infty(2B_\alpha)} +O(\alpha).
\eea
\end{theorem}
\proof
Equalities  \eqref{i21}-\eqref{i22} imply

\bea
\Psi_{0,a}(\frac{\nabla H_{0,a}(z^\star)^\perp}
{| \nabla H_{0,a}(z^\star)|}) &=& \min_{\xi\in \partial
  B}\Psi_{0,a}(\xi) \\&=&
\min_{\xi\in \partial
  B}\left(\Psi_{0,b}(\xi)+\Psi_{0,a}(\xi)-\Psi_{0,b}(\xi)\right). 
\eea
Therefore 
\bea
\min_{\xi\in \partial
  B}\left(\Psi_{0,b}(\xi)-|\Psi_{0,a}(\xi)-\Psi_{0,b}(\xi)\right)|
 \leq \Psi_{0,a}(\frac{\nabla H_{0,a}(z^\star)^\perp}{| \nabla 
 H_{0,a}(z^\star)|})\leq \min_{\xi\in \partial
  B}\left(\Psi_{0,b}(\xi)+|\Psi_{0,a}(\xi)-\Psi_{0,b}(\xi)|\right),
\eea
which implies 
\bean \label{jj1}
|\Psi_{0,a}(\frac{\nabla H_{0,a}(z^\star)^\perp}
{| \nabla H_{0,a}(z^\star)|})-\Psi_{0,b}(\frac{\nabla H_{0,b}(z^\star)^\perp}
{| \nabla H_{0,a}(z^\star)|}) | \leq \max_{\xi\in \partial
  B}|\Psi_{0,a}-\Psi_{0,b}|,
\eean

and consequently

\bea
|\Psi_{0,a}(\frac{\nabla H_{0,a}(z^\star)^\perp}
{| \nabla H_{0,a}(z^\star)|})-\Psi_{0,a}(\frac{\nabla H_{0,b}(z^\star)^\perp}
{| \nabla H_{0,a}(z^\star)|}) | \leq 2\max_{\xi\in \partial
  B}|\Psi_{0,a}-\Psi_{0,b}|,
\eea
Using the explicit expression of $\Psi_{0, a}(\xi)$ given in
\eqref{psi0}, we find
\bea
\left| \frac{\nabla H_{0,a}(z^\star)^\perp }{| \nabla
    H_{0,a}(z^\star)|} - \frac{\nabla H_{0,b}(z^\star)^\perp}
{| \nabla H_{0,b}(z^\star)|}  \right| \leq C \max_{\xi\in \partial
 B}|\Psi_{0,a}-\Psi_{0,b}|. 
\eea  
Since  $\varepsilon_{s,a}(z^\star) $ is lower
bounded, 
Combining the estimate above and  \eqref{jj1}, we obtain the 
desired result. 

\endproof

Now, by combining the results of theorems \eqref{resul} (corollary
\eqref{main2bis}), and   \eqref{main3}, we have the main stability estimate
in theorem \eqref{main}.

%%%%%%%%%%%%%%%%%%%%%%%%%%%%%%%%%%%%%%%
\section{Appendix}
%%%%%%%%%%%%%%%%%%%%%%%%%%%%%%%%%%%%%%%
In this section we derive the asymptotic  expansion of the gradient 
of the Green function 
$\nabla_x G(z^\star+\alpha \xi, z^\star+\alpha \xi^\prime)$ when $\alpha$ tends to zero. 
%%%%%%%%%%%%%%%%%%%%%%%%%%%%%%
\begin{theorem}\label{app1}
Let $G(x, y)$ be the Green function solution to the system \eqref{Green1}-
\eqref{Green2}. Then, the following asymptotic expansion holds
\bea
\alpha \partial_{x_k}G(z^\star+\alpha \xi, z^\star+\alpha
\xi^\prime) &=& 
\varepsilon_s(z^\star)\partial_{\xi_k}\Phi_0(\xi, \xi^\prime) +
\frac{1}{4\pi} \partial_{x_k}\varepsilon_s(z^\star) \alpha \ln(\alpha)
+\alpha \Phi_1(\xi, \xi^\prime) +o(\alpha),
\eea
for all $\xi, \xi^\prime \in B$ satisfying  $\xi\not=\xi^\prime$, and
$o(\alpha)$  is uniform in $\xi, \xi^\prime \in B$.  \\

$\Phi_0(\xi,\xi^\prime) = \frac{1}{2\pi}\ln(|\xi-\xi^\prime|)$
is the Green function of the Laplacian in the whole space,   and  
$\Phi_1(\xi,\xi^\prime) $ has a logarithmic singularity on 
the diagonal $\xi= \xi^\prime$, that is $ |\Phi_1(\xi,\xi^\prime)|
\leq C |\Phi_0(\xi,\xi^\prime)|$, for  all  $\xi, \xi^\prime \in B$,
with $C>0$ is constant that only depends on $\varepsilon_s(x)$.

\end{theorem}
%%%%%%%%%%%%%%%%%
\proof 
We first use the Liouville transformation and substitute the Green 
function $G(x,y)$ by 
\bea
\underline{G}(x,y) = \frac{1}{
\varepsilon_s^{\frac{1}{2}}(x)
\varepsilon_s^{\frac{1}{2}}(y)}
G(x,y),
\eea
in the system \eqref{Green1}-
\eqref{Green2}, to obtain
\begin{eqnarray}  \label{Gr1}
\Delta \underline{G}(x,y)
+V(x)
\underline{G}(x,y) &= 
\delta_{y}(x)&
\quad \textrm{in} \quad \mathbb R^2.
\end{eqnarray} 
with the  Sommerfeld radiation condition as $|x| \rightarrow +\infty$:
\bean \label{Gr2}
\frac{\partial \underline G}{\partial |x|} -i
\omega\sqrt{\varepsilon_0 \mu_0}\underline G&=& O(\frac{1}{\sqrt{|x|}}),
\eean
and where 
\bean \label{Valpha}
V(x)&:= &\omega^2 \mu_0 \varepsilon_s(x)
-\frac{\Delta \varepsilon_s^{\frac{1}{2}}(x)}
{\varepsilon_s^{\frac{1}{2}}(x)}.
\eean
For simplicity, we assume that $V(z^\star) \not=0$. If it is not
the case the proof can be slightly modified. \\

Let $\underline G_0(x, y) $ be the Green function of the
Helmholtz
equation in the free space, solution to the system
\begin{eqnarray}  \label{Gr01}
\Delta \underline{G}_0(x, y)
+V(y)
\underline{G}_0(x, y) &= 
\delta_{y}(x)&
\quad \textrm{in} \quad \mathbb R^2.
\end{eqnarray} 
with the  Sommerfeld radiation condition as $|x| \rightarrow +\infty$:
\bean \label{Gr02}
\frac{\partial \underline G_0}{\partial |x|} -i
\sqrt{V(y)}\underline{G}_0&=& 
O(\frac{1}{\sqrt{|x|}}).
\eean

The function $\underline G_0(x, y) $ is given by
\bea
\underline G_0(x, y) = -\frac{i}{4}H_0^{(1)}
(\sqrt{V(y)}|x-y|), \qquad  
\textrm{for  } x\not= y, 
\eea
where $H_0^{(1)}(t)$ is the Hankel function of the first kind of order
zero.\\

Now, we shall derive the asymtotic expansion of
$\partial_{x_k}\underline G(x,y) $ as $x$ tends to $y$. 

Let \bea
\mathcal G(\xi, \xi^\prime) :=  
\underline G(x, y)-
\underline G_0(x, y).\eea
 
It satisfies the Helmholtz equation 
\begin{eqnarray}  \label{Gr1}
\Delta \mathcal G (x, y)
+V(x) \mathcal G(x, y)&= 
 -(V(x)-V(y))\underline{G}_0(x, y)&
\quad \textrm{in} \quad B_R(z^\star).
\end{eqnarray} 
with the boundary condition 
\bean \label{Gr2}
\mathcal G (x, y) = \underline G(x, y)-
\underline G_0(x, y) \quad \textrm{on} 
\quad \partial B_R(z^\star). 
\eean
Further we fix $R>1$ such that the system \eqref{Gr1}-\eqref{Gr2}
has a unique solution. Since the  $H_0^{(1)}(t)$ has a 
logarithmic singularity as $t$ tends to zero, the right hand side belongs 
to $C^{0, \iota}( \overline{B_R(z^\star)})$ for any $\iota\in [0,1)$,
uniformly in $y \in B_1(z^\star)$ (see for instance Proposition 4.1 in 
\cite{BT}).\\

 Considering the fact that $\underline G(x, y)-
\underline G_0(x, y) \in C^\infty(\partial B_R(z^\star)\times
B_1(z^\star))$, we deduce from elliptic regularity   that  
$\mathcal G (x, y) \in C^{2, \iota}( \overline{B_R(z^\star)} )$
uniformly  in  $y \in B_1(z^\star)$ \cite{McLean}. In addition, due to
the explicit expression of the  right hand side in equation
\eqref{Gr1}, on can prove easily that 
$\partial_{x_k}\mathcal G(z^\star+\alpha \xi, z^\star+\alpha
\xi^\prime)$ has a finite continuous limit when $\alpha$ tends to zero, denoted 
by $\Phi_{11}(\xi, \xi^\prime)$. \\

From known asymptotic expansions of Hankel functions, we have ~ \cite{Abra}
\bea
\partial_{x_k}\underline G_0(z^\star+\alpha \xi, z^\star+\alpha
\xi^\prime) = \frac{1}{\alpha} \partial_{x_k}\Phi_0(\xi, \xi^\prime)
+ \alpha\ln(\alpha)|\xi-\xi^\prime|+O(\alpha), 
\eea
where $O(\alpha)$ is uniform in  $\xi, \xi^\prime \in B$.  \\

Consequently 
\bea
\alpha \partial_{x_k}\underline G (z^\star+\alpha \xi, z^\star+\alpha
\xi^\prime) =  \partial_{x_k}\Phi_0(\xi, \xi^\prime)
+ \alpha\Phi_{11}(\xi, \xi^\prime)  +o(\alpha), 
\eea
which combined with the regularity 
of $\varepsilon_s(x)$ achieves the proof of the theorem.

\endproof

%%%%%%%%%%%%%%%%%%%%%%%%%%%%%%%%%%%%%%%%%%%%%%%

%%%%%%%%%%%%%%%%%%%%%%%%%%%%%%%%%%%%%%%%%%%%%%%%%%%%%%

\end{document}